\documentclass{PoS}

\title{Spin Physics: session summary}

\ShortTitle{Spin Physics: session summary}

\author{Christine Aidala\\
        Los Alamos National Laboratory\\
        E-mail: \email{caidala@bnl.gov}}

\author{Simonetta Liuti\\
        University of Virginia\\
        E-mail: \email{sl4y@cms.mail.virginia.edu}}

\author{
{Caroline Riedl}%
        \thanks{A footnote may follow.}\\
       DESY\\
       E-mail: \email{criedl@mail.desy.de}}

\abstract{We summarize the main results of the Spin Physics Working Group Sessions at DIS 2010, the XVIII International Workshop on Deep Inelastic Scattering and Related Subjects.}

\FullConference{XVIII International Workshop on Deep-Inelastic Scattering and Related Subjects\\
                April 19 -23, 2010\\
                Convitto della Calza, Firenze, Italy}

\begin{document}

\section{Introduction}
\label{sec:intro}

The year 2010 marks the start of data taking at the LHC. This threshold has been accompanied by an increased awareness of the importance of the underlying hadronic physics components which need to be known to great accuracy. In particular, Parton Distributions Functions (PDFs), and more recently, Generalized Parton Distributions Functions (GPDs) along with  Transverse Momentum Distributions (TMDs) will be necessary elements for extracting and better interpreting a number of the collider's observables. Viceversa, the new experimental data from the LHC will contribute to understanding unsolved theoretical issues on PDFs, GPDs and TMDs in untested regimes.

\begin{figure}
\label{fig1}
\centerline{\includegraphics[width=0.45\textwidth]{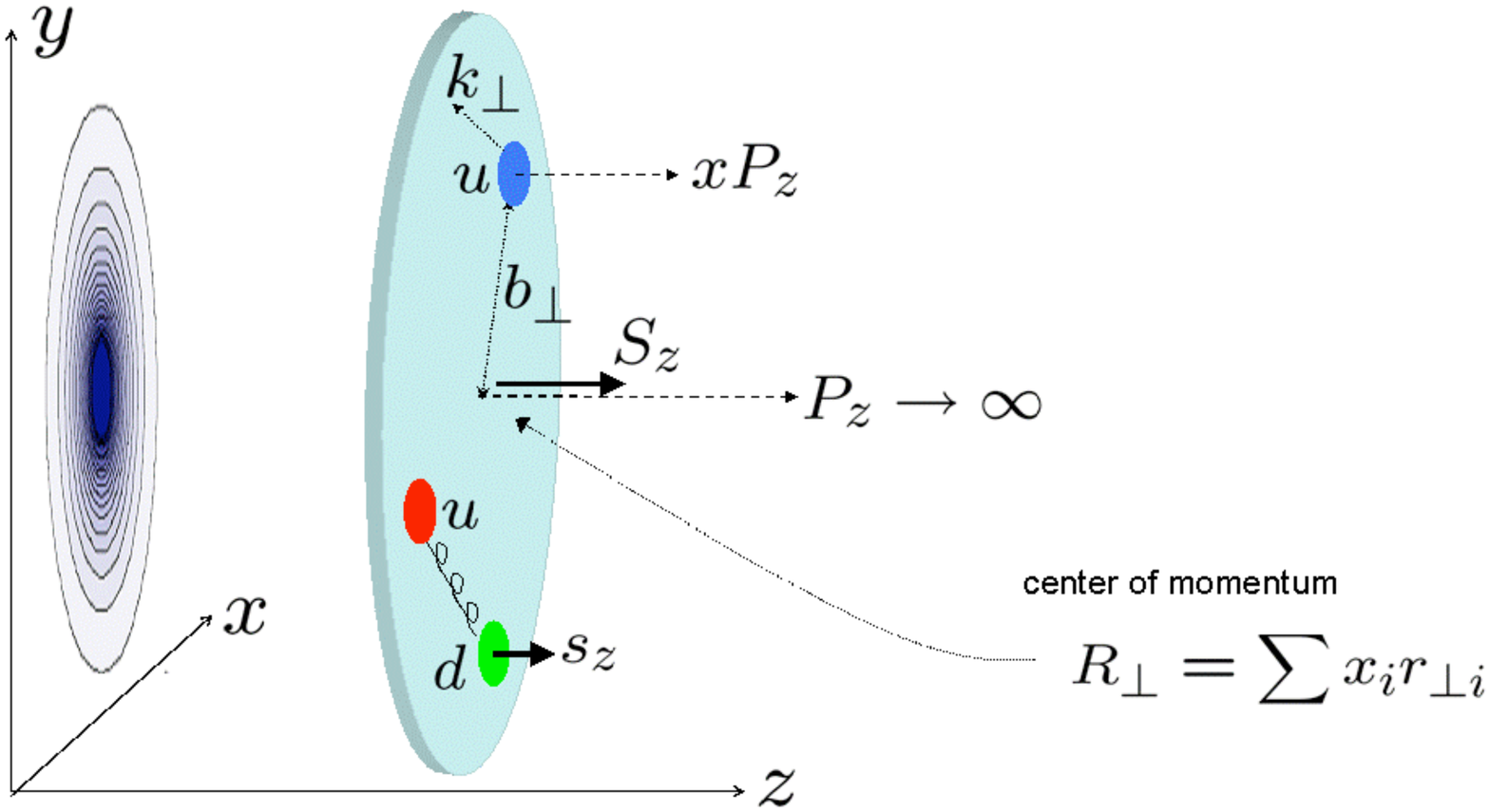}}
\caption{A sketch of the proton's structure as obtained from  both longitudinal and transverse polarization observables \cite{Hagler}. }
\end{figure}

GPDs and TMDs are rather newly measurable observables obtained from a variety of experiments such as deeply virtual exclusive scattering, Drell-Yan processes, Semi-inclusive Deep Inelastic Scattering (SIDIS),  with polarized beams and/or targets. A large and diverse set of experimental data is being, and will be produced, from which these important components of hadronic structure need to be extracted and compared to theoretical predictions.  Experimental observations will lead to a great improvement of the spin, spatial, and momentum correlations in hadronic structure (a pictorial representation is given in Fig.\ref{fig1}). Extensive programs to measure polarized PDFs, GPDs, and TMDs have been held and are currently in place at several laboratories around the world including HERMES, Jefferson Lab, CERN/Compass, Brookhaven National Laboratory. GPDs and TMDs measurements represent also an important component in the planning of the future Electron Ion Collider (EIC).

The importance of spin observables such as GPDs and TMDs is, in a nutshell, that they provide us with ``tomographies" of hadrons (Fig.\ref{fig1}). 
On one side, TMDs allow us to explore within QCD the deformations of the partons'  transverse momentum distributions due to the different polarization states of both the beam and the target. On the other, GPDs provide a unique access to the partons' transverse spatial distributions. 
This type of information is obtained by examining the various helicity structures in SIDIS, and in a variety of exclusive scattering observables. 
It is a goal that was made only recently achievable thanks to the increased capability of present and future experiments that  will afford us the possibility of performing such delicate analyses.
In this contribution we give an assessment of the most recent developments in spin physics, both the theoretical issues, and the up to date measurements obtained from deeply virtual scattering processes, at the inception of the LHC.

From the theoretical point of view, we single out a few new trends. More refined analyses are starting to appear  concerning  developments on artificial neural network type studies of polarized observables. These developments follow the recent realization that both hadronic systems and hadronic collisions  can be seen as multi-particle events described in terms of an  enlarged number of kinematical variables, and  evolving from a varied set of initial conditions.

At the same time more focused studies on the partonic interpretation of GPDs are appearing. GPDs do not have an obvious partonic interpretation in the so-called ERBL kinematical region where the returning quark momentum takes on negative values. This situation has been interpreted as given by a quark-antiquark pair emerging from the proton. However, it is by now established that its concrete realization within QCD requires multi-parton components/reinteractions. This has been obtained so far only within a class of models using Light Front wave functions. In addition, an observation was made recently that because of the dispersion relations obeyed by the exclusive amplitudes,  all the information on GPDs might be concentrated on the kinematical ridge between the ERBL and the more standardly interpreted DGLAP region -- where all quark's momenta are positive. While inferring partonic relations from dispersion relations for exclusive amplitudes needs to be taken with a grain of salt due to the $t$-channel impact on the kinematical thresholds,  we can conclude that a fully consistent partonic interpretation is still lacking, and a new role for partonic reinteractions at leading order might be key to the solution of this puzzle. We also report on the progress that is continuously made on outstanding issues which were started over the past few years.  These include the modeling of the Orbital Angular Momentum (OAM) contribution to the spin sum rule, the working of factorization for unintegrated parton distributions leading to the development of new QCD evolution equations, and finally the inclusion of error analyses in the spin dependent parametrizations.

On the experimental side, we will highlight recent measurements performed at the spin facilities at {\sc BNL}, {\sc Cern}, {\sc Desy}, {\sc JLab}, and {\sc KEK}. The measured helicity distributions (see Sec.~\ref{sec:heli}) serve as input for more and  more improved global fits, putting recently a better constraint especially on $\Delta G$, the gluon spin contribution.  Semi-inclusive measurements probing  TMDs (see Sec.~\ref{sec:tmds}) provide not only evidence for non-vanishing orbital angular momentum of valence quarks, but moreover allow a statement about its sign, while exclusive measurements related to GPDs (see Sec.~\ref{sec:gpds}) allow in principle a constraint on the total angular momentum of quarks.
These theoretical frameworks provide two complementary approaches to an aspect of spin physics that has in recent years often been denoted 'Nucleon Hologram'. On the exclusive sector, the measurements at different facilities complement each other with regard to the covered kinematic phase space and the observables.

In what follows we summarize and discuss both the theoretical and experimental progress in the field as it appears from the Spin Working Group contributions. Both theoretical and experimental presentations were arranged in several subgroups:
\begin{itemize}
\item Helicity Structure of the Nucleon from Inclusive Reactions \cite{Zemlyanickina,Chiu,Balewski,Surrow,Manion,Franco,Korotkov,deJager}
\item SIDIS and Transverse Spin \cite{Schnell,Martin,Rossi,Vossen,Gordon,Dharmawardane,Ruiz,Savin,Boer,Zavada,Kang,Ceccopieri,Cisbani,Qiu,Lyuboshitz}
\item Exclusive Processes \cite{Goldstein,Marukyan,Teryaev,Girod,Taneja,Mahon,Meyer}
\end{itemize}
A session on Parametrizations of PDFs and TMDs \cite{Blumlein,Rojo,Perry,Melis}, and joint sessions with the Diffraction \cite{Augustyniak,Gorshteyn,Wallon},
 and Future Facilities groups were also organized \cite{Prokudin,Horn,Diehl,Klasen}.

The world-wide aspect of spin physics makes, on the one hand, our subject a very exciting one, on the other hand it had quite a significant impact on the course of the DIS 2010. Only a few days prior to the start of the conference, an Iceland volcano with the beautiful name Eyjafjallaj\"okull decided to send a cloud of ashes into the European air space. This made not only the European air traffic collapse, but also caused the cancellation of many transatlantic flights. We gained a lot of experience not only with the European train system (one of us traveled by train from Glasgow to Florence), but also with the EVO conferencing system. 

We wish to thank the extraordinary efforts of all the attendees, who managed to be with us either traveling amidst the many cancellations, or connecting via EVO sometimes even at odd hours. A major positive outcome of this conference was in witnessing the incredible enthusiasm, persistence, and dedication that we all share in our field of research.

\section{Helicity Structure of the Nucleon}
\label{sec:heli}

One main focus of Deep Inelastic Scattering (DIS) lies in the task of unraveling the spin content of the nucleon, whose spin quantum number is 1/2. The benchmark spin experiments in the last two decades of the 20th century revealed that the nucleon's spin is not solely made up from the spins of its quarks,
$\Delta\Sigma\equiv\int_0^1\mathrm{d}x\sum_{q,\overline{q}}\left[\Delta q(x)+\Delta\overline{q}(x)\right]$.
Other candidates to solve the 'spin puzzle' are the spin of the gluons, $\Delta G$, and the orbital angular momentum $L$ of quarks and of gluons.

\begin{figure}[t]
\begin{tabular}{c}
\includegraphics[width=.50\textwidth]{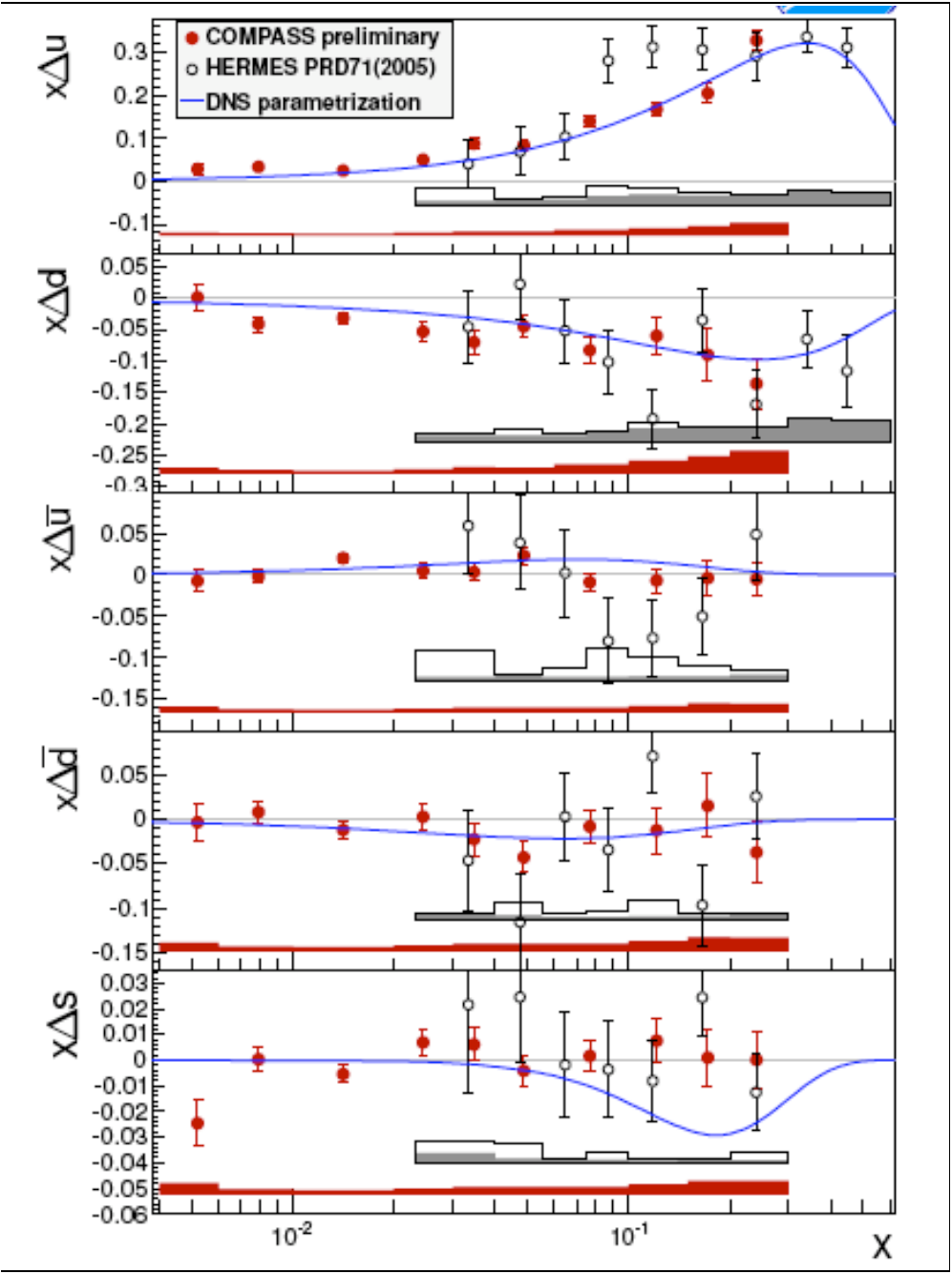}
\end{tabular}
\begin{tabular}{c}
\includegraphics[width=.36\textwidth]{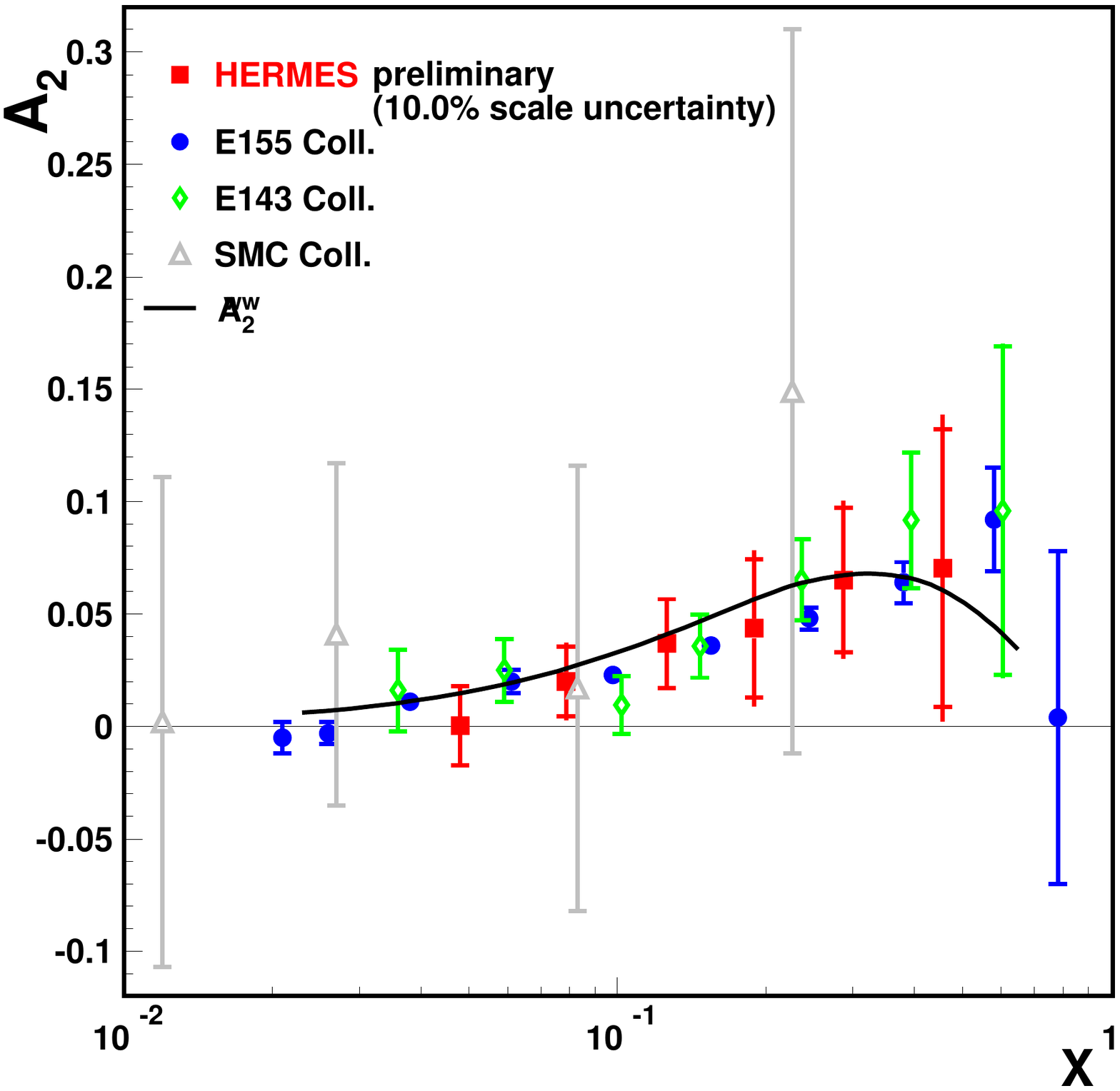}\\
\includegraphics[width=.36\textwidth]{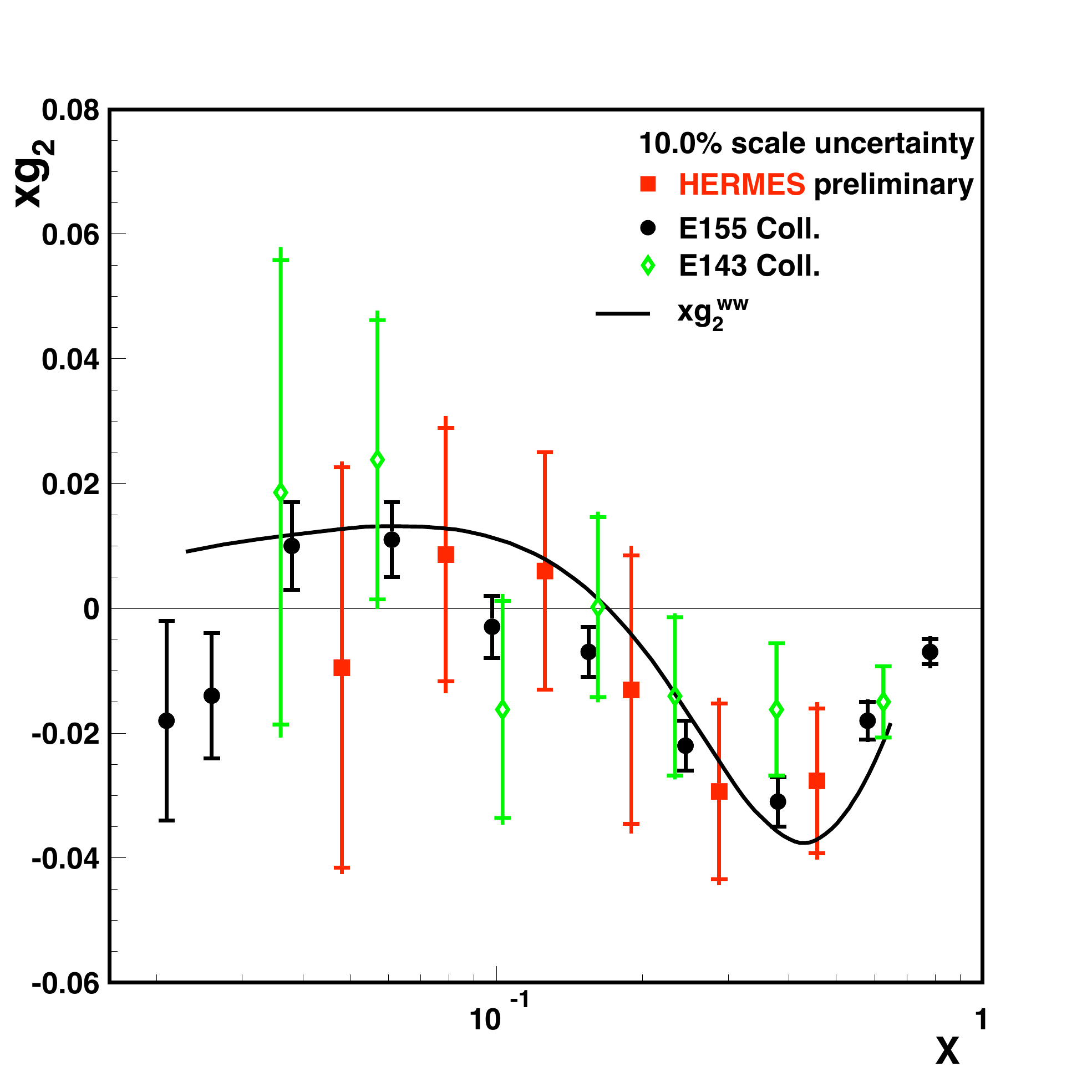}
\end{tabular}
\caption{Left: the spin-dependent, flavor-separated parton distribution functions vs. $x$-Bjorken, as extracted by {\sc Compass} from proton- and deuteron-target data. Right: (top) the longitudinal double-spin asymmetry $A_2$ and (bottom) the inclusive structure function $g_2$ measured by {\sc Hermes}.}
\label{fig:compass-pdfs}
\end{figure}

{\sc Compass} showed preliminary results on the polarized PDFs $x\Delta q(x)$, $q=u,d,\overline{u},\overline{d},s$ \cite{Zemlyanickina} extracted from all 2002-2006 published deuteron data and the 2007 preliminary proton data. New longitudinal proton data are expected for 2011. The SIDIS analysis allows for a flavor separation of the double-spin asymmetries for different hadron species, from which the polarized PDFs were obtained by using the known parameterizations of the unpolarized PDFs and the fragmentation functions at a scale of 3 GeV$^2$. The results are in good agreement with the {\sc Hermes} results (see Fig.~\ref{fig:compass-pdfs}). All sea quark distributions are found to be compatible with zero. The non-strange PDFs are in good agreement with results of previous global QCD fits, while the shape of $\Delta s(x)$ disagrees significantly with these fits. A small flavor asymmetry of the light sea quarks is observed, with $\Delta\overline{u}\gtrsim\Delta\overline{d}$. Moreover, the data allow a successful test of the Bjorken sum rule.
{\sc Compass} also presented the longitudinal spin-transfer from the fragmenting quark to a $\Lambda$-baryon for data taken with longitudinally polarized beam and proton target \cite{Kang}. Through its self-analyzing weak decay $\Lambda\rightarrow p\pi^-$, the $\Lambda$ baryon is an ideal probe to study spin effects. Apart from probing the spin-transfer mechanism, this measurement allows for a test of the spin-[in]dependent asymmetry of the strange quark sea $\Delta s(x)\neq\Delta\overline{s}(x)$ [$s(x)\neq\overline{s}(x)$]. The longitudinal spin-transfer was found to be positive and rising with $x_F$ for the $\overline{\Lambda}$, but to be compatible with zero for the $\Lambda$. This finding is not in agreement with the new preliminary {\sc Hermes} result \cite{Rith}, which suggests a positive value of about 0.2 for the spin-transfer to the $\Lambda$.

\begin{figure}[t]
\centerline{
\includegraphics[width=.50\textwidth, angle=-90]{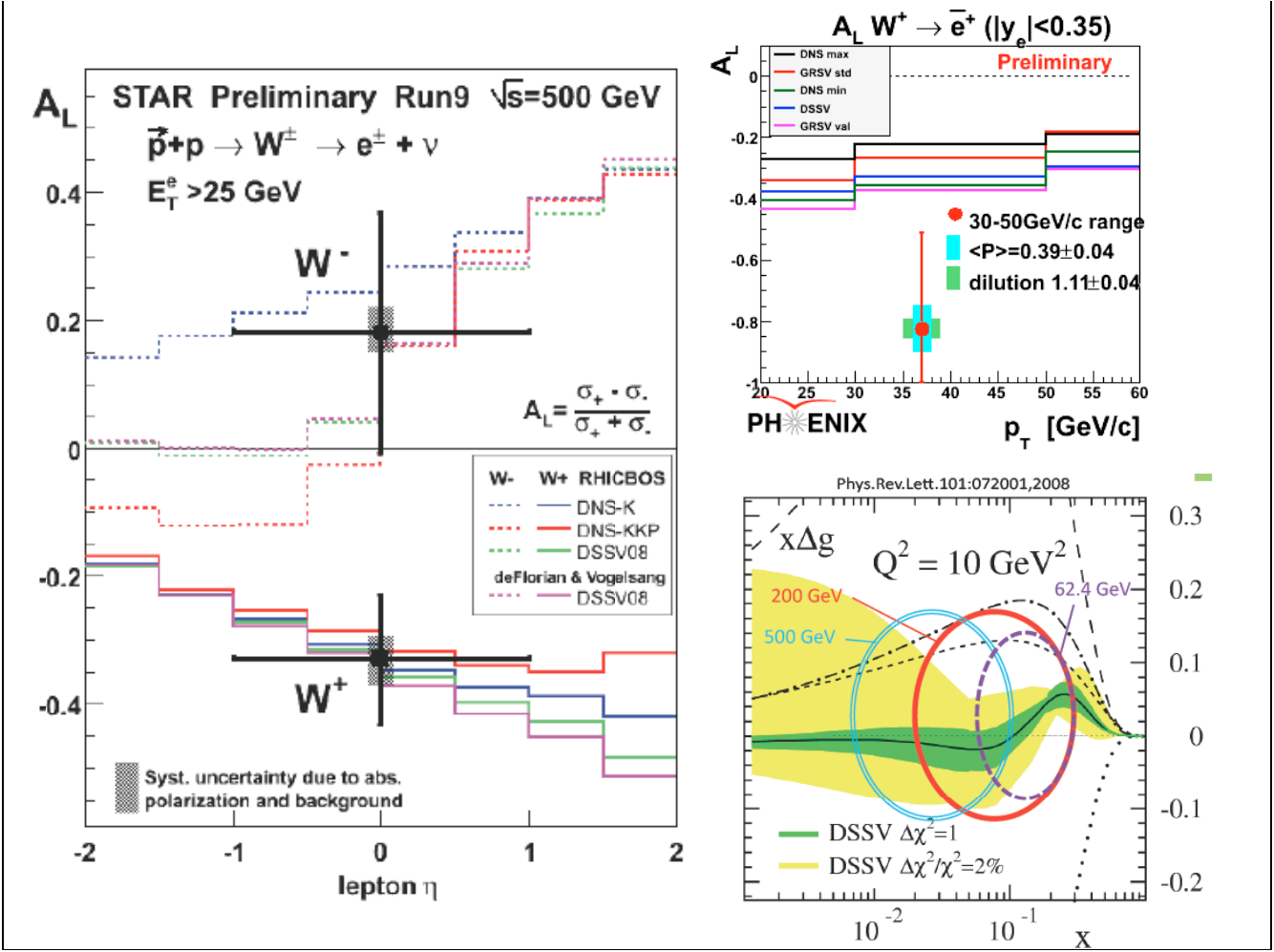}
}
\caption{Recent spin results from {\sc Rhic}: single-spin asymmetry in $W$ production and gluon polarization.}
\label{fig:bnl-heli}
\end{figure}

Shown in the left and upper-right panels of Fig.~\ref{fig:bnl-heli} are preliminary results from the {\sc Star}~\cite{Balewski} and {\sc Phenix}~\cite{Chiu} experiments at {\sc Rhic} for the single-longitudinal (parity-violating) spin asymmetry in $W$ production.  The results are based on a few weeks of $\sqrt{s}=500$~GeV $p+p$ commissioning data taken in 2009 and mark the beginning of a new phase in the {\sc Rhic} spin program, as 500~GeV running is expected to be the focus for the next several years.  Through the $W$ boson's direct coupling to flavor, these measurements provide access to the flavor-separated sea quark helicity distributions and are complementary to existing semi-inclusive DIS measurements, with no reliance on fragmentation functions and at a much higher scale ($Q^2 \approx m_W^2$).  These initial measurements have large uncertainties but do show good agreement with predictions based on fits to (semi-inclusive) DIS data, and the $W^+$ measurements are already several sigma from zero, confirming the large parity-violating asymmetry expected.

With polarized proton-proton collisions, the {\sc RHIC} program also offers leading-order access to gluons, and there were reports on the ongoing gluon helicity program from both {\sc Phenix}~\cite{Manion} and {\sc Star}~\cite{Surrow}.  The 2008 parametrization of the gluon helicity distribution by de~Florian, Sassot, Stratmann, and Vogelsang \cite{DSSV}, which includes neutral pion data from {\sc Phenix} at $\sqrt{s} = 200$~GeV and 62.4~GeV, and jet data from {\sc Star} at 200~GeV, in addition to world DIS and SIDIS data, can be seen in the lower-right panel of Fig.~\ref{fig:bnl-heli}.  The circled regions illustrate how the $x$ range probed at {\sc Rhic} can be varied by changing the center-of-mass energy.

{\sc Compass} showed a new result of the gluon polarization in the nucleon obtained from $D^0$ double-spin asymmetries in photon-gluon fusion tagged by open-charm meson production \cite{Franco}. In addition to the 2002-2006 data, for the first time the 2007 data were included in the analysis. A neural network was used to model the signal and background distributions. $\Delta G/G$ was extracted in leading order and was found to be $-0.08\pm0.21(\pm0.11)$ at $\langle x_G\rangle =0.11$ and a scale $\langle\mu^2\rangle=13$ GeV$^2$. A next-to-leader analysis of these data is under study. The world data on $\Delta G$ of fixed-target facilities indicate that all experimental measurements are in agreement with each other, showing a preference for small values of $\Delta G$. Note that all these measurements are afflicted with large experimental uncertainties and only cover a restricted range in $x_G$ of at most $\langle x_G\rangle=0.03...0.4$. For the first time, the final {\sc Hermes} result on $\Delta G$ \cite{Hermes-DeltaG} was included in the world data overview plot.

One of the four structure functions needed to describe the nucleon in inclusive DIS is $g_2(x,Q^2)$, which 
is related to an integral of the spin structure function $g_1(x,Q^2)$ ({\sc Wandzura-Wilczek} relation). The sensitivity of the inclusive cross-section to $g_2$ reaches the maximum under the experimental condition that the angle between the lepton scattering plane and the plane containing the incoming lepton and the target polarization vector is 90$^\circ$. {\sc Hermes} presented a preliminary analysis of the double-spin asymmetry $A_2$ and the spin structure function $g_2$ \cite{Korotkov} including the complete data set collected on the transversely polarized proton target in 2002-2005. The data were subject to a QED radiative and detector smearing unfolding. The {\sc Hermes} data are within the covered kinematic range in good agreement with older measurements at {\sc Slac} and {\sc Smc} and provide a successful test of the {\sc Wandzura-Wilczek} relation (see Fig.~\ref{fig:compass-pdfs}). In order to recover the full statistical power of the {\sc Hermes} data, the covariance matrix arising from the kinematic unfolding must be taken into account for world-data fits and integrals.

The most important theoretical advancements have been in the determination of the gluon contribution to the proton spin sum rule, $\Delta G$, either indirectly through perturbative QCD evolution, or directly, through the channels which are sensitive to the gluon component.   While pQCD evolution is known precisely to NLO,  an important issue is the uncertainty or bias in the various parametric forms for the polarized quark distributions. A very careful analysis was presented by Blumlein \cite{Blumlein}, that includes an evaluation of both the experimental and theoretical systematic uncertainties, along with an extraction of the HTs in the entire kinematical region.  Preliminary results by Rojo \cite{Rojo} address the question of the bias inherent in the choice of given parametric forms, using the NNPDF approach. Nonetheless the results still show large error bars for the gluon distribution $\Delta g$, and they cannot predict whether this is
is a positive or negative quantity. 
In a similar spirit, a new approach using Self-Organizing Maps (SOMs) was discussed by Perry \cite{Perry}. 
SOMs are a type of neural network whose nodes form a topologically ordered map during the learning process.
The learning process is {\it unsupervised}  at variance with standard NNs.

\section{Spin-momentum Structure of the Nucleon and TMDs}
\label{sec:tmds}

Transverse Momentum Dependent PDFs, or TMDs, view the nucleon not only in terms of the longitudinal momentum fraction but also the transverse momentum carried by its partons, thereby encoding the correlation of this momentum and the spin of the parton and/or nucleon. 



\begin{figure}[h!]
\begin{tabular}{c}
\includegraphics[width=.48\textwidth]{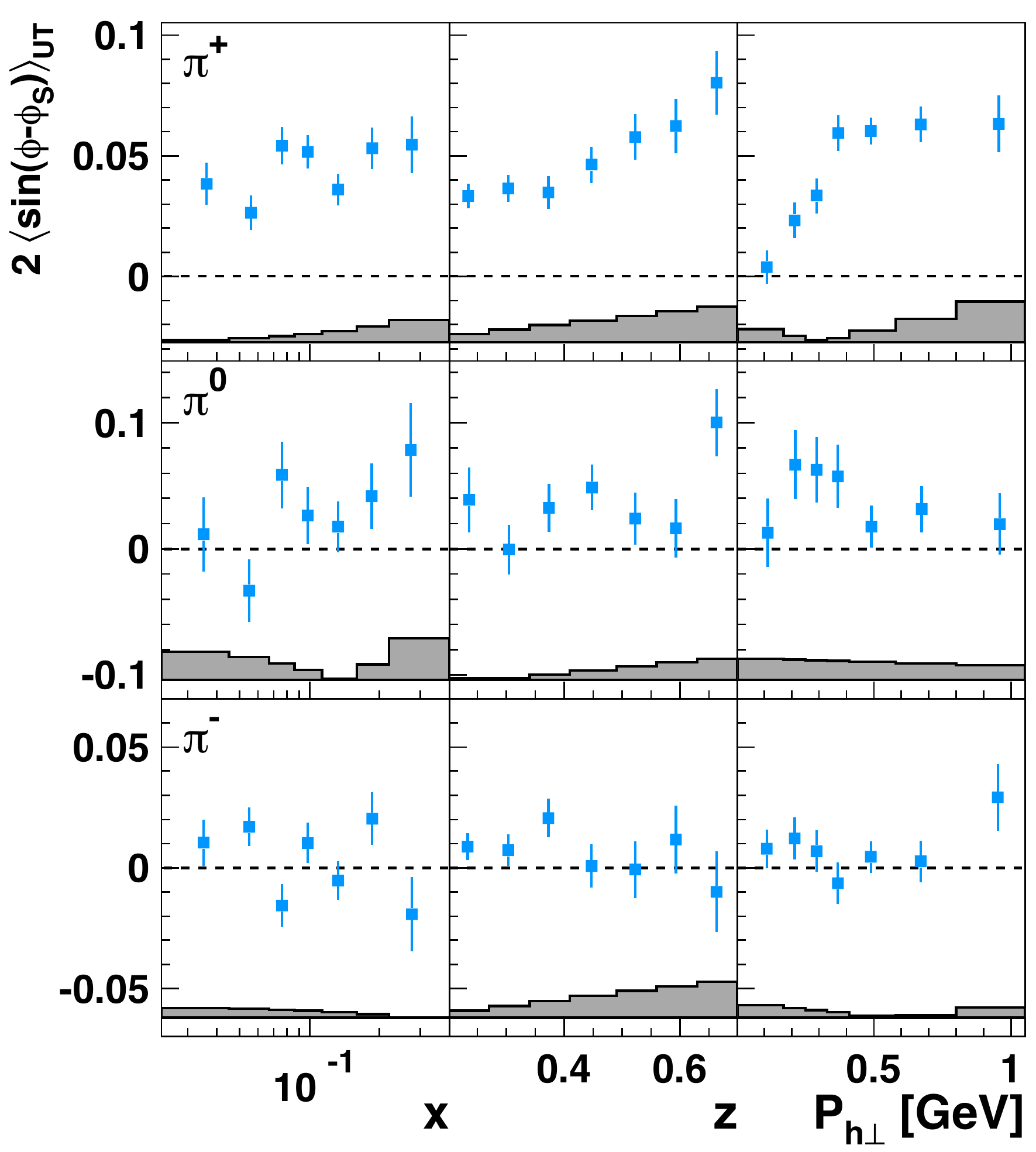}
\end{tabular}
\begin{tabular}{c}
\includegraphics[width=.48\textwidth]{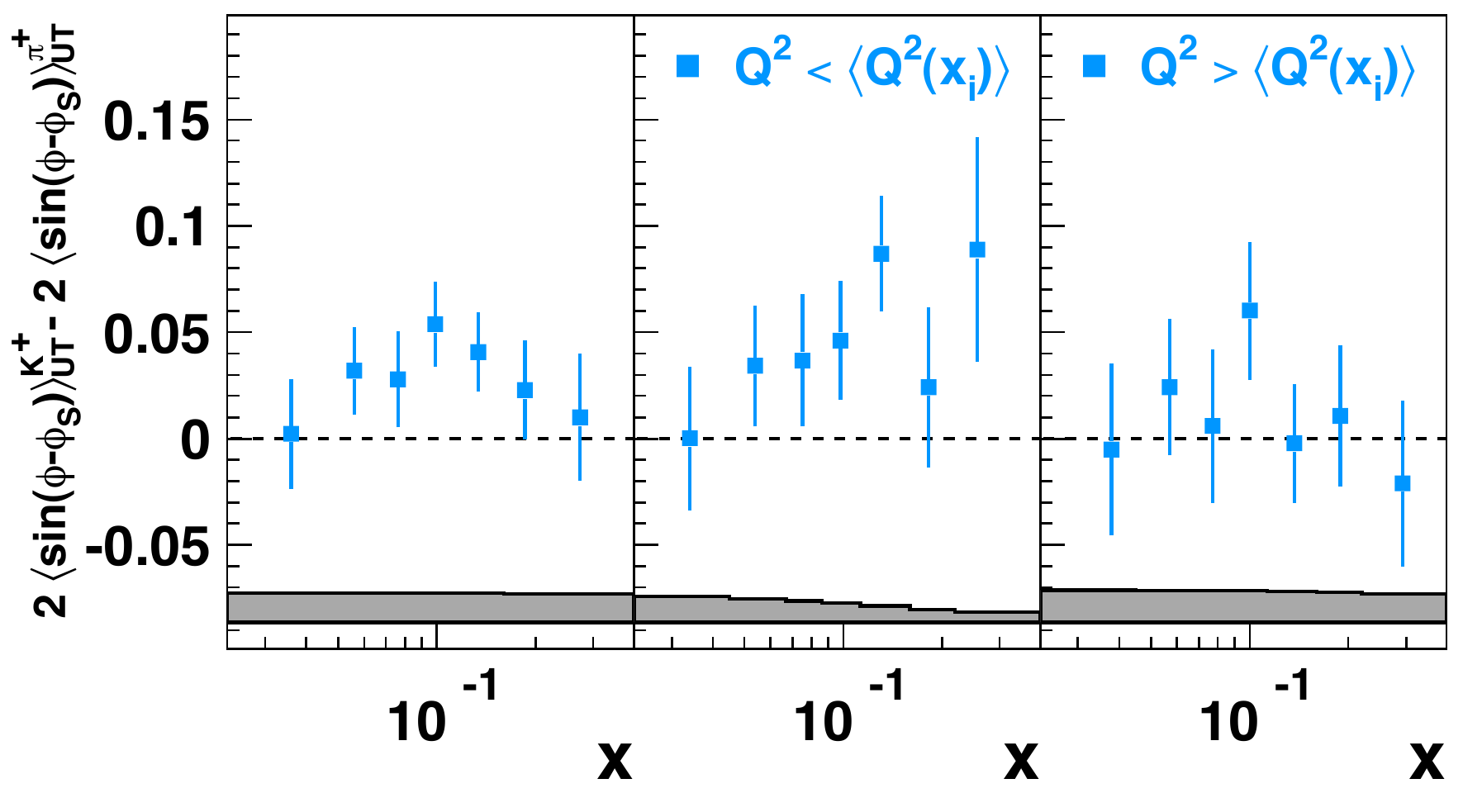}\\
\end{tabular}
\centerline{
\includegraphics[width=.48\textwidth, angle=-90]{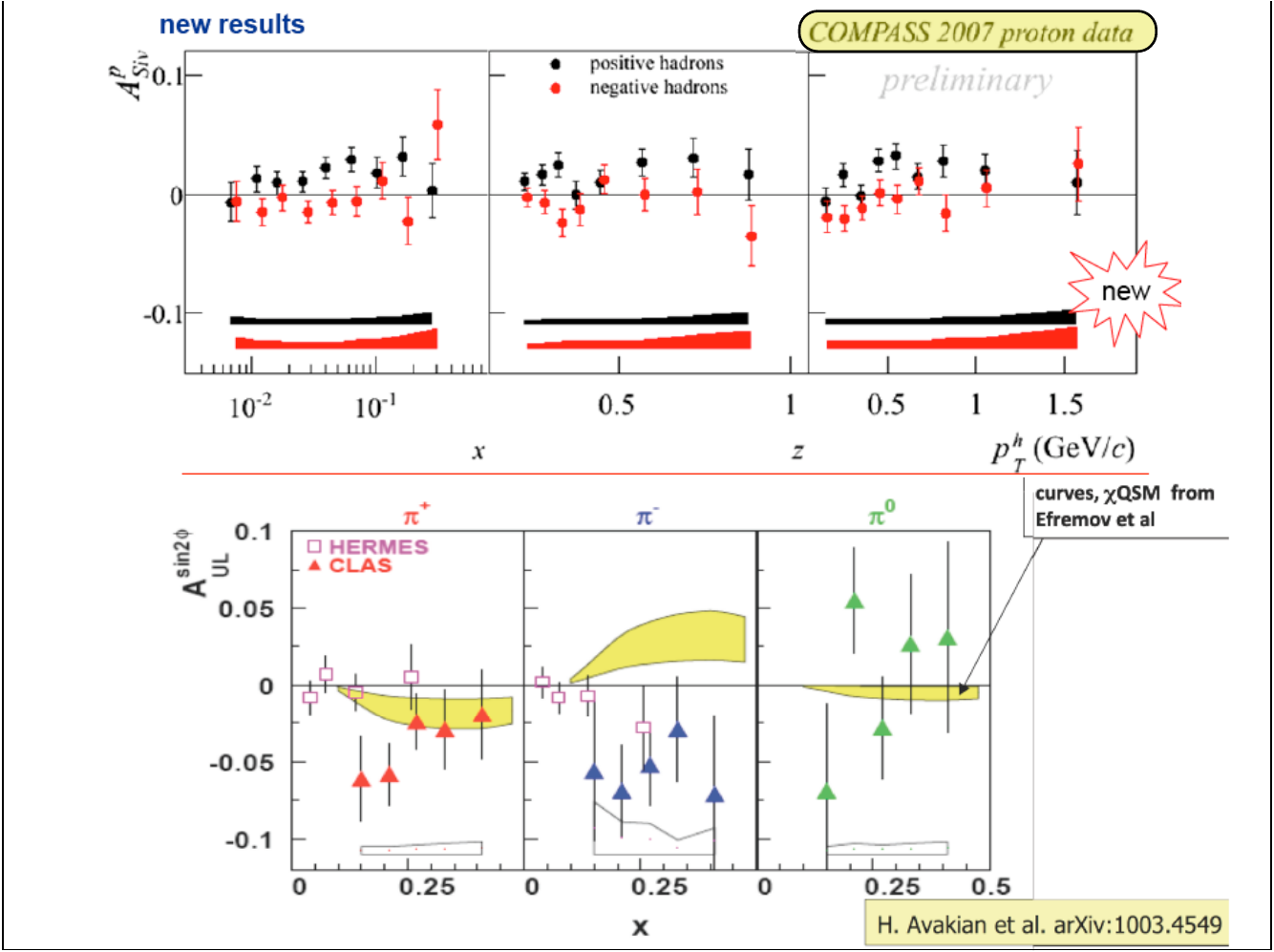}
}
\caption{Top: {\sc Hermes} Sivers amplitudes for identified pions. Left figure, from top to bottom for positive, neutral, and negative pions, in projection vs.~$x$, $z$, and $P_{h\perp}$. 
Right hand side panels: the Sivers amplitudes of positive kaons minus that of positive pions vs.~$x$, for all $Q^2$ values, and separated into the low and high $Q^2$ region.
Bottom: {\sc Compass} Sivers amplitudes for positive and negative hadrons, and the {\sc Clas} worm-gear TMD related measurement of single-spin amplitudes for identified semi-inclusive pions off a longitudinally polarized target. }
\label{fig:hermes-tmds}
\end{figure}

The various harmonic modulations of the SIDIS cross-section in one-hadron production on a transversely polarized target are related to different TMDs. The relevant azimuthal angles $\phi$ and $\phi_S$ are defined in Fig.~\ref{fig:hermes-tmds}.
{\sc Hermes} presented final results of the Sivers $\sin(\phi-\phi_S)$ amplitudes on a proton target and preliminary results of other modulations \cite{Schnell}. The Sivers amplitudes for positively charged pions are clearly positive and rising with $z$ and $P_{h\perp}$, reaching a plateau at $P_{h\perp}\simeq 0.4$ GeV, see Fig.~\ref{fig:hermes-tmds}, while the amplitudes for neutral pions are slightly positive and those for negative pions compatible with zero. Not shown in the figure is the amplitude for positive kaons, which is positive, and for negative kaons, which is slightly positive. Due to the dominance of u-quark scattering for the production of positive pions, these findings suggest a large and negative Sivers function for u-quarks, while the vanishing amplitudes for negative pions require cancellation effects, e.~g.~a d-quark Sivers function of the opposite sign.
The Sivers effect provides evidence for the orbital motion of quarks inside a transversely polarized nucleon: the quark's angular motion generates a left-right density difference, causing the u-quarks to favor ``left'' and the d-quarks ``right''.
Also shown in Fig.~\ref{fig:hermes-tmds} is the Sivers amplitude difference between positive kaons and positive pions, which is slightly positive, thereby indicating possible contributions from sea quarks. In the case of pure u-dominance, this difference should be zero.  Similar evidence for sea quark contributions has been observed by the {\sc Brahms} experiment at {\sc Rhic}, which observed positive transverse single-spin asymmetries of the same magnitude for both positive and negative kaons, as well as a surprising non-zero asymmetry in anti-proton production but proton asymmetry consistent with zero.  {\sc Hermes} found no significant non-zero signal for the $\sin(3\phi-\phi_S)$ modulation related to the so-called pretzelosity TMD. 
This finding indicates that pretzelosity is either very small, or that its suppression level by powers of $P_{h\perp}$ in SIDIS is too high. Also no signal was found for the sub-leading twist $\sin(2\phi+\phi_S)$ and $\sin(2\phi-\phi_S)$ amplitudes, both related to the so-called worm-gear TMD describing transversely polarized quarks in a longitudinally polarized nucleon (or vice versa). A significant non-zero signal for negatively charged mesons is observed for the sub-leading $\sin(\phi_S)$ amplitude, of which various terms are related to the transversity, worm-gear, or Sivers TMDs. The final {\sc Hermes} results of the Collins fragmentation function, which correlates the transverse polarization of a quark and the transverse momentum of the produced hadron, were published very recently.

{\sc Compass} showed new TMD results extracted from the 2007 transversely polarized proton data \cite{Martin}, providing an extended kinematic domain compared to {\sc Hermes}. While the Sivers amplitude for negatively charged unidentified hadrons is consistent with zero, these new Run-2 data give evidence for a positive signal for positively charged hadrons, however smaller by a factor of 2 than the {\sc Hermes} results. This might be due to an unexpected $W$-dependence of the Sivers effect, explaining the difference between the results of the two experiments by different average values of $W$ at the same average $x$; however, the physical significance of a possible $W$ dependence remains unclear.
The previously published {\sc Compass} deuteron data
show no Sivers signal. Together with {\sc Hermes} data, this further supports a large and positive d-quark Sivers function. For the {\sc Compass} Collins asymmetry on the proton target, a clear positive signal is found in the valence region, in agreement with the {\sc Hermes} result. This non-zero signal provides evidence that both the Collins fragmentation function and the transversity distribution are non-zero. The Collins asymmetry compares well to a global fit including data from {\sc Hermes}, {\sc Compass} and {\sc Belle} \cite{Anselmino}. The previously published Collins asymmetry on the deuteron target was found to be zero, which is interpreted as an effect of cancellation of u- and d-quark transversity distributions. Other harmonic modulations in the SIDIS cross-section are under analysis, and a {\sc Compass} high-luminosity run on a transversely polarized proton target has started in May 2010.

{\sc Clas} will take data on a transversely polarized HD-ice target in 2011. There are also plans for transverse running after the {\sc JLab12} upgrade. Currently, TMD-related information is extracted from data with an unpolarized or longitudinally polarized target \cite{Rossi,Harut}. {\sc Clas} results on the target-spin asymmetry in semi-inclusive production of pions were shown. For a longitudinally polarized target, the only modulation arising at leading order is the $\sin(2\phi)$-moment, which is related to the worm-gear (also called Kotzinian-Mulders) TMD. For charged pions, a non-zero negative $\sin(2\phi)$-moment was found, while this moment is found to be compatible with zero for neutral pions.
A similar analysis was presented by {\sc Compass} for the semi-inclusive production of unidentified charged hadrons off a longitudinally polarized deuteron target \cite{Savin}. All harmonic modulations $\sin(n\phi)$ were found to be compatible with zero for positive integer $n$.
{\sc Clas} also presented the longitudinal double-spin asymmetry with respect to beam and target for semi-inclusive neutral and charged pions. The $p_T$-dependence can be interpreted in terms of different TMD-widths resulting from the different orbital motion of quarks polarized in the direction of the proton spin, and opposite to it.
Preliminary {\sc Clas} results of the beam-spin asymmetry for pions were shown. The $\sin\phi$-moments for neutral and positively charged pions are clearly positive and compatible with each other, indicating a small Collins-type contribution for positive pions. The results are in the kinematic overlap region in agreement with the {\sc Hermes} measurement.

Projections for the Sivers- and Collins-moment of the single-spin asymmetry in semi-inclusive production of charged pions on a transversely polarized Helium-3 target measured in Hall A at {\sc JLab} were shown \cite{Cisbani}, representing the first TMD measurement on a neutron target. Between the conference and the submission of these proceedings, the results have released to the public, including the first measurement of the double-spin asymmetry for a longitudinally polarized beam and a transversely polarized target.

{\sc Belle} presented the first measurement of chiral-odd, collinear interference fragmentation functions \cite{Vossen}, which describe the fragmentation of a transversely polarized quark into two spin-zero hadrons and which carry an azimuthal $\sin\phi$ modulation. The fragmentation functions can therefore be measured via azimuthal asymmetries of di-hadron correlations in e$^+$e$^-$-annihilation into quarks. The extracted preliminary asymmetry is of significant size. The future goal is a combined analysis of SIDIS, $p+p$, and e$^+$e$^-$-data in order to extract the transversity distribution and to disentangle the contributions to the transverse single-spin asymmetry $A_N$.

The transverse-momentum-dependent Sivers distribution function discussed above also plays an important role in transverse single-spin asymmetries in $p+p$-collisions and was in fact first proposed to describe large (up to $\sim 40$~\%) transverse single-spin asymmetries observed in polarized $p+p$ scattering in the 1970s and '80s.  As presented by {\sc Star}~\cite{Gordon} and {\sc Phenix}~\cite{Dharmawardane}, these inclusive asymmetries, which may be due to multiple contributing effects, persist in pion production up to {\sc Rhic} energies, $\sqrt{s} = 200$~GeV, and contrary to predictions that the asymmetry should decrease with pion transverse momentum as $\frac{1}{p_T}$, the asymmetries measured by both experiments are consistent with rising or flat up to $p_T \sim $3--4~GeV/$c$.  Another surprise from {\sc Rhic} that remains unexplained is that {\sc Star} has found evidence that the asymmetry in inclusive $\eta$ meson production is larger than that of neutral pions, which could possibly be due to strangeness, mass, or isospin effects.  While interpretation of inclusive single-spin asymmetries in $p+p$ collisions remains challenging due to the number of possible contributing effects, as more is learned from the simpler systems of $e^+ + e^-$ and DIS, more can in turn be learned from the $p+p$ data.  As we probe more deeply into this rapidly developing field of QCD dynamics in hadrons, hadron-hadron collisions will play a critical role, of great interest at present e.g. to test the predicted process dependence of T-odd TMDs such as the Sivers function, and to investigate the predicted TMD factorization breaking in processes involving more than two hadrons \cite{Rogers}.

As discussed above, in $p^\uparrow + p$-experiments of the past three decades, several significantly non-zero left-right asymmetries $A_N$ steeply increasing with $x_F$ were observed in inclusive hadron production, with opposite sign for different hadron charges. One explanation is the Sivers effect suggesting different left-right preferences for u- and d-quarks.
The study of this asymmetry in $ep^\uparrow$-scattering can serve as clean test of the TMD formalism and as test whether the asymmetry goes as predicted to zero at high values of $p_T$.
{\sc Hermes} presented the first measurement of the transverse single-spin asymmetry in inclusive hadron production in  $ep^\uparrow$-scattering \cite{Ruiz}. The $\sin\phi$-amplitude of the asymmetry is clearly positive for positively charged pions and kaons, rising with $p_T$ until about $p_T\equiv 0.8$ GeV and then decreasing. For negative pions and kaons, the asymmetry-amplitude is significantly smaller, which can be understood as fingerprint of u-quark dominance in the proton. This behavior versus $p_T$ resembles, apart from the decrease, the Sivers effect, whose amplitude reaches a plateau for higher values of the transverse momentum of the produced positive pion (as shown in Fig.\ref{fig:hermes-tmds}).

{\sc Compass} presented preliminary results on transverse $\Lambda$ polarization in $\Lambda$-production off a transversely polarized proton target \cite{Kang}, which allows access to transversity. For the $\Lambda$, the transverse polarization is found to be compatible with zero, while the data suggest a small negative value for the $\overline{\Lambda}$ with no prominent $x$- or $z$-dependence in either case.

Theoretical talks focused on the extraction and parametrization of 
the transversity distribution, $h_1$, which measures the quarks polarization transverse to the direction of motion, the various structure functions measuring  
spin-transverse momentum correlations at leading twist -- the Sivers function, $f_{1T}^\perp$, the Collins function, the Boer-Mulders function, $h_{1T}^\perp$-- and the twist-4 Cahn contribution sensitive to the unpolarized transverse momentum dependence. 
Talks by Melis \cite{Melis} and Prokudin \cite{Prokudin} presented state of the art parametrizations obtained from increasingly sophisticated global analyses. An important point is that in the parametric forms of \cite{Melis,Prokudin} the $k_T$ dependence factorizes, a feature that might be important for the QCD interpretation of the correlator, and that has 
been confirmed within errors in lattice calculations \cite{Hagler}.
In \cite{Prokudin}  the parametrizations were used to emphasize the importance 
of a larger kinematical coverage attainable at the planned  high luminosity Electron Ion Collider.   

The possibility of extracting and interpreting universal, long-distance  functions through global fits to experiments hinges
on QCD factorization theorems. Two complementary approaches for SIDIS were discussed at the workshop: TMD Factorization, based on the original introduction of leading order Final State Interaction (FSI) effects in Ref.\cite{BroHwaSch}, and Collinear Factorization (CF). Qiu \cite{Qiu} presented an improved analysis in the CF scheme that considers twist-3 correlation functions. In this talk a future plan was also outlined including developments on the
pQCD evolution of the twist-3 functions.
Another important related question is the $k_T$ evolution. In \cite{Ceccopieri} it was shown that 
in SIDIS and Drell Yan processes one can apply the evolution for parton jets development  including the two sources of $k_T$, from the non perturbative fragmentation and from jet evolution.

Finally, new studies of $\Lambda$ polarization  in unpolarized scattering experiments are emerging and were presented at the workshop \cite{Boer}. Surprising
results showing large polarizations in hyperons produced from high energy unpolarized protons scattering, first appeared in the pioneering measurements at BNL (for a review see \cite{Panagio}). They can be considered yet another facet of the spin puzzle. QCD-based mechanisms have been invoked to explain the generation of hyperons' transverse  polarization. Boer's approach, which extends beyond collinear factorization, introduced another degree of freedom in the fragmentation stage, the so-called polarizing fragmenation function, $D^\perp_{1T}$.

\section{Nucleon Hologram and GPDs}
\label{sec:gpds}

Generalized Parton Distributions (GPDs) embed at the same time information on the longitudinal momenta of partons (like Parton Distribution Functions, PDFs) and their transverse position inside the nucleon (as usually probed in measurements of elastic form factors).
GPDs can be used to describe the soft, non-perturbative part of hard exclusive reactions, in which the final state is entirely resolved. At leading twist (i.~e.~twist 2) and for a spin-1/2 target such as the proton, there are four GPDs needed for chiral-even processes, i.~e.~processes that do not flip the chirality of the involved quark. Another four GPDs that are chiral-odd appear in processes related to transversity, which were discussed in Sec.~\ref{sec:tmds}.

The cleanest channel of such exclusive reactions is Deeply Virtual Compton Scattering (DVCS), or the hard electroproduction of a real photon, where the target stays intact and a quark in the nucleon emits a single real photon. Other channels, such as exclusive vector meson production, involve an additional meson amplitude. The observables are cross-sections and asymmetries with respect to the azimuthal angle between the lepton scattering plane and the plane spanned by the recoiling target and the produced photon or meson. For a transversely polarized target, there is a second azimuthal angle defined by the target-polarization vector (as depicted in Fig.~\ref{fig:hermes-tmds} for one-hadron production).
In the analysis of the data, the measured asymmetries are subject to a harmonic expansion with respect to the azimuthal angle(s). The extracted Fourier coefficients (or asymmetry amplitudes) are related to the real or imaginary parts of a combination of the complex Compton form factors (CFFs) $\mathcal{H},\mathcal{E},\mathcal{\widetilde{H}},\mathcal{\widetilde{E}}$ \cite{BKM}, which are convolutions of the corresponding twist-2 GPDs $H,E,\widetilde{H},\widetilde{E}$ with the hard scattering kernel.

In the case of DVCS, there is a second process with the same initial and final state, in which the real photon is emitted from the initial or scattered charged lepton ({\sc Bethe-Heitler}, BH) and which therefore interferes with the DVCS process. This interference allows for DVCS-sensitive measurements also at experiments for which the DVCS process is rather sub-dominant compared to BH.

\begin{figure}[t]
\begin{tabular}{c}
\includegraphics[width=.48\textwidth]{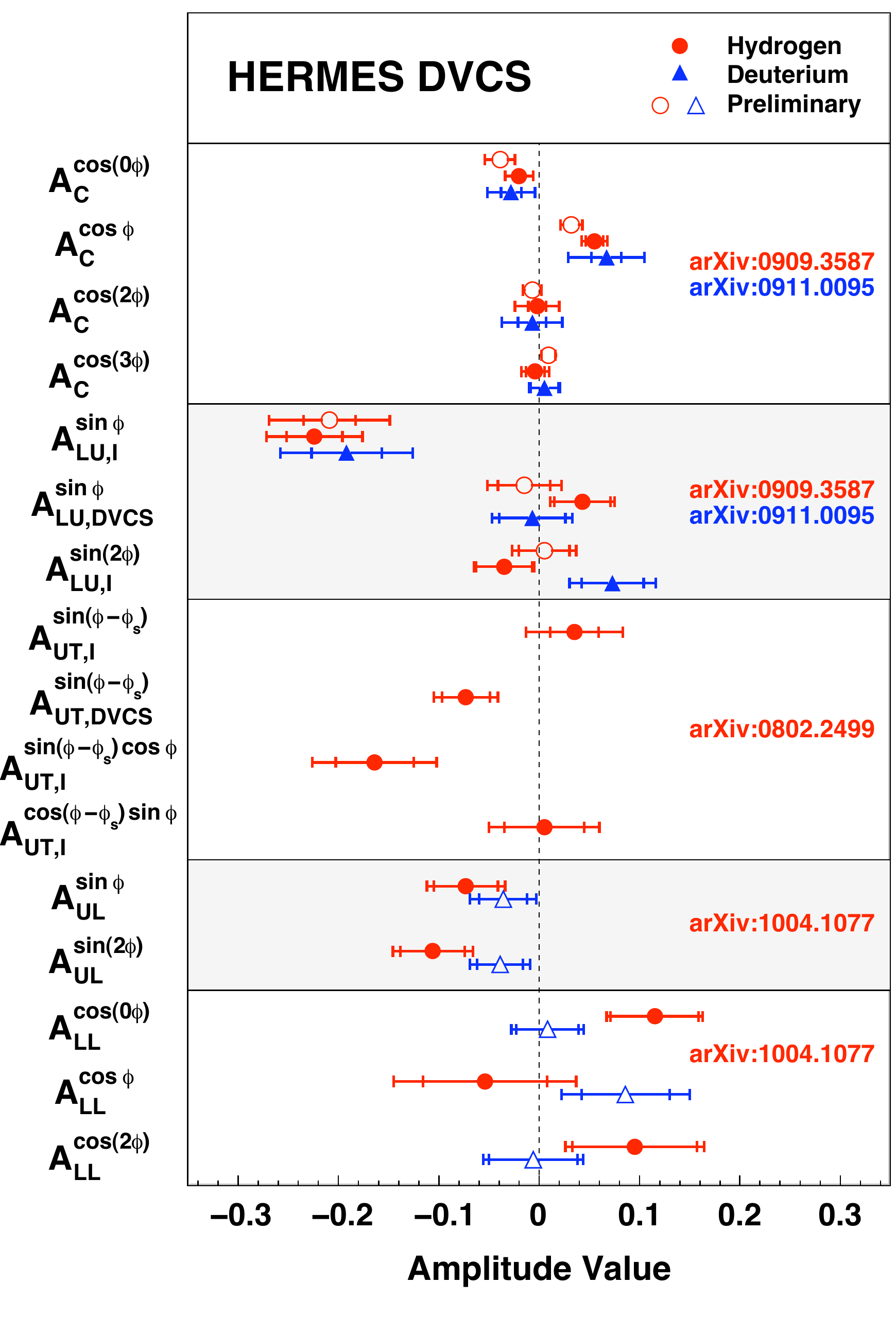}
\end{tabular}
\begin{tabular}{c}
\includegraphics[width=.48\textwidth]{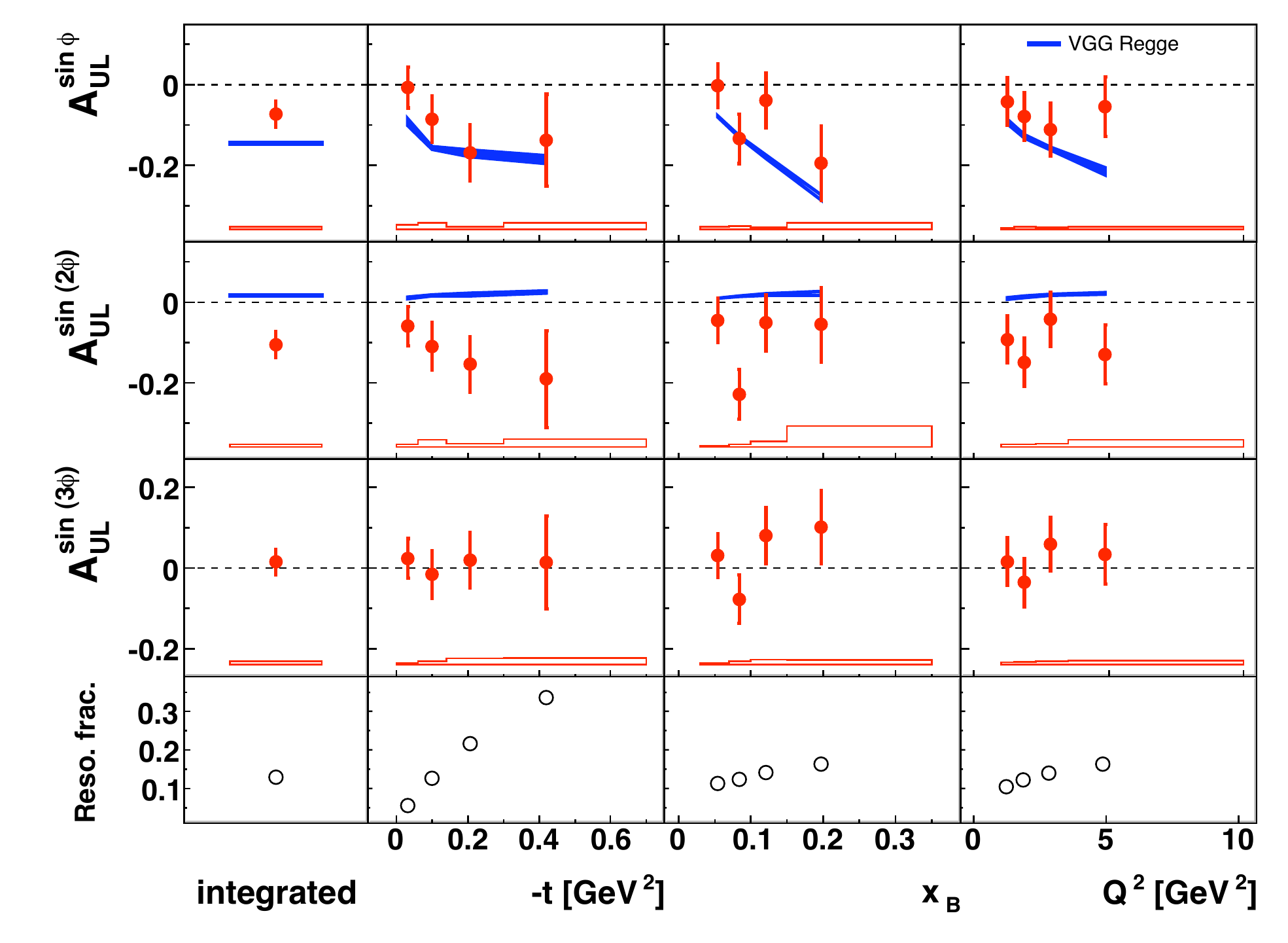}\\
\includegraphics[width=.48\textwidth]{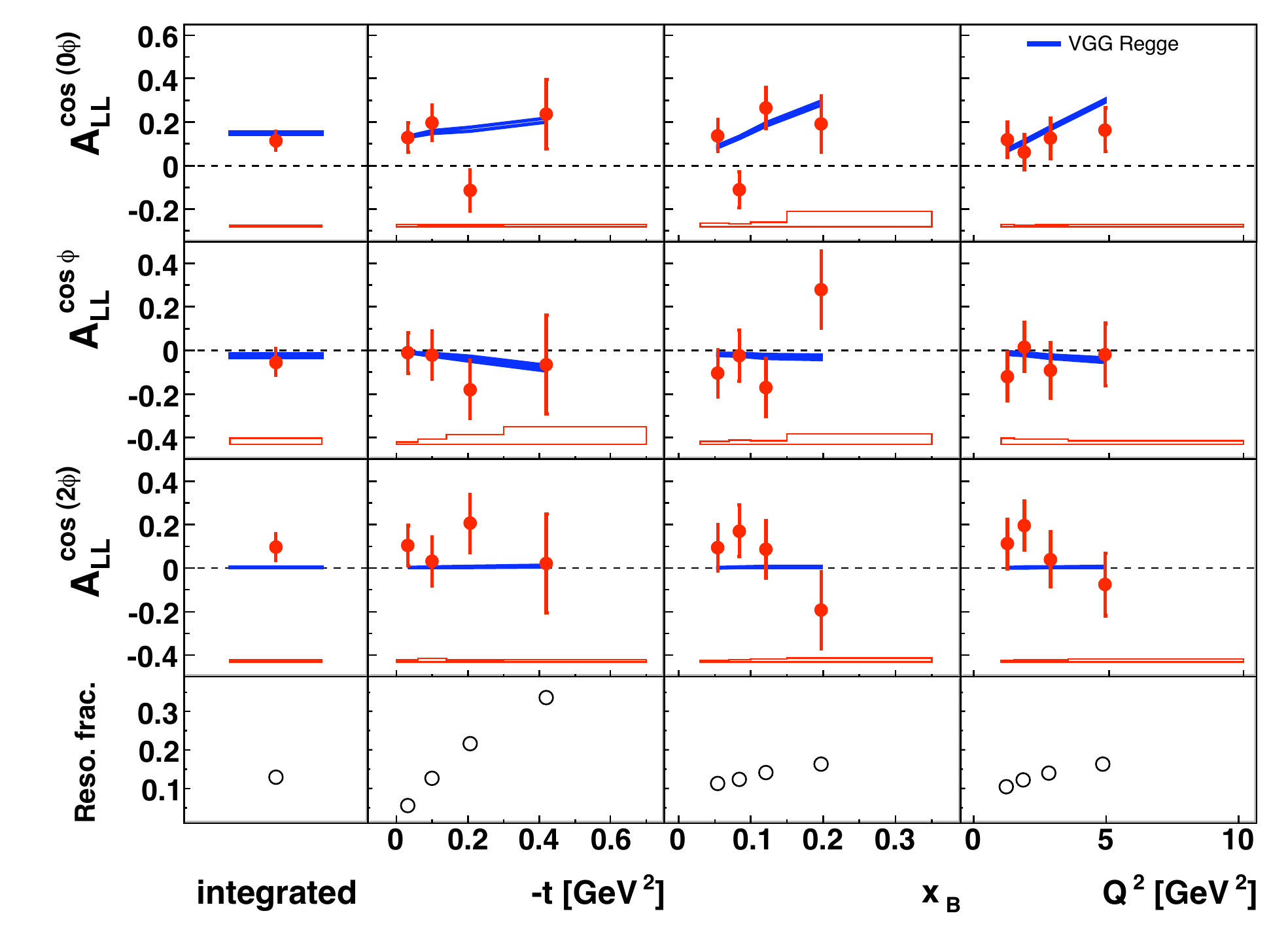}
\end{tabular}
\caption{Azimuthal asymmetry amplitudes measured by {\sc Hermes} in deeply-virtual Compton scattering on a hydrogen (red circles) and deuterium target (blue triangles). The left panel shows the results extracted over the entire available kinematic range with average values of $\langle Q^2\rangle=2.5$ GeV$^2$, $\langle x_B\rangle=0.10$, and $\langle -t \rangle=0.12$ GeV$^2$. 
The right panels show the kinematic projections versus $-t$, $x_B$, and $Q^2$ of the single- (top) and double-spin (bottom) asymmetry amplitudes \cite{Marukyan,Air_GPDs}. 
}
\label{fig:hermes-dvcs}
\end{figure}

{\sc Hermes} presented the final results on beam-helicity- and beam-charge asymmetries in DVCS on unpolarized hydrogen and deuterium targets (1996-2005 data set) and for the first time preliminary high-statistics hydrogen results from the 2006/2007 data \cite{Marukyan,Air_GPDs}. Because {\sc Hera} provided two lepton beam charges, the leading sinusoidal amplitudes from the squared DVCS-term and the interference term can be separated. While the beam-charge asymmetry projects out the real part of CFF $\mathcal{H}$, the (charge-difference) beam-helicity asymmetry probes the imaginary part of CFF $\mathcal{H}$. When the target is polarized, the extracted DVCS asymmetry amplitudes are sensitive to other GPDs. A compilation of all {\sc Hermes} DVCS measurements on hydrogen and deuterium targets is shown in Fig.~\ref{fig:hermes-dvcs} (left) for the average kinematics. All data are also available in projections versus $-t$, $Q^2$, and $x_B$, as shown in the right part of the figure for the very recent publication of the single- and double-spin asymmetries in DVCS on a longitudinally polarized proton target \cite{Mahon}. The target-unpolarized hydrogen data were also analyzed in a two-dimensional $(x_B,-t)$-binning. The measurements on the deuterium target offer the interesting possibility of probing differences between spin-1/2 (nucleon) and spin-1 (deuteron) GPDs. Nine GPDs are needed in the chiral-even leading-twist case to describe a spin-1 hadron. At low values of the squared momentum transfer $t$ to the target, DVCS is assumed to happen on the deuteron (coherent scattering) and not on a nucleon inside the deuteron. For the polarized deuterium data, there is the additional CFF $\mathcal{H}_5$ related to the inclusive tensor structure function of the deuteron in the forward limit. Within the experimental uncertainties, no significant difference was found between hydrogen and deuterium data in the region $-t<0.06$ GeV$^2$ with about 40$\%$ coherent contribution, i.~e.~neither a clear coherent, nor tensor signature could be confirmed. The target-unpolarized data set 2006/2007 including the information from a Recoil detector is still under analysis. With the detection of the recoiling target proton, the elastic signal can be separated from the contribution of associated processes, in which the target proton is excited to a resonant state.

\begin{figure}[t]
\centerline{
\includegraphics[width=.50\textwidth, angle=-90]{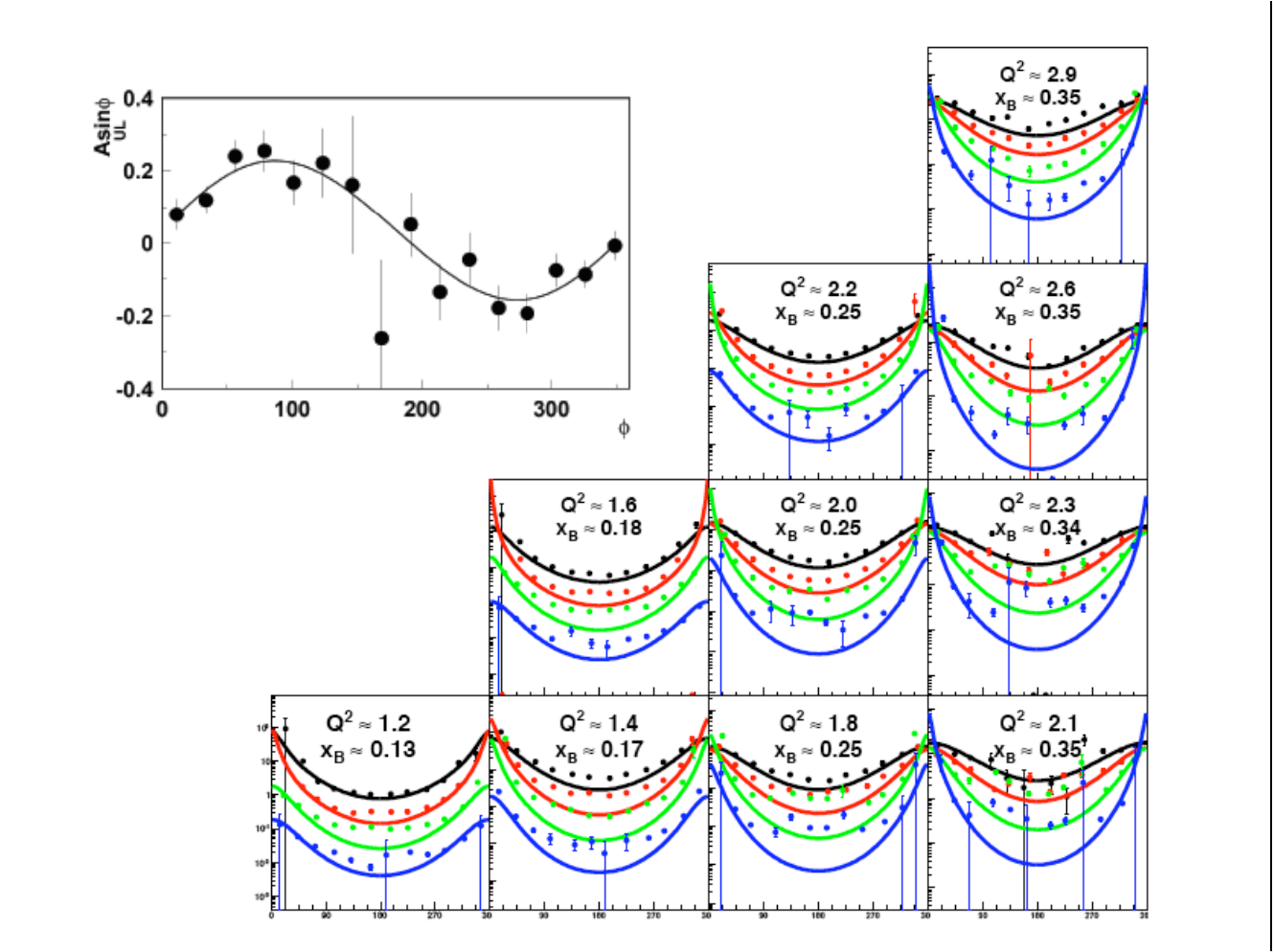}
}
\caption{Left upper panel: {\sc Clas} longitudinal target-spin asymmetry in DVCS versus the azimuthal angle, showing a sinusoidal modulation and superimposed with a harmonic fit. Only about 20$\%$ of the accumulated statistics are shown. The other panels display the DVCS cross-section in different $(x_B,Q^2)$-bins. Furthermore, in each of the little panels, different $-t$-bins are indicated by different colors, together with the calculated BH-contribution.}
\label{fig:clas-dvcs}
\end{figure}

{\sc Clas} presented several DVCS measurements \cite{Girod,Harut}. The {\sc Clas} kinematic average of $\langle Q^2\rangle=1.8$ GeV$^2$, $\langle x_B\rangle=0.28$, and $\langle -t \rangle=0.31$ GeV$^2$ is different from {\sc Hermes}, but the two coverages have some overlap. Shown were the single-charge beam-spin asymmetry published in 2008 and the preliminary single- and double-spin asymmetry results extracted from a dedicated high-statistics run in 2009 with a newly designed inner calorimeter and longitudinal target polarization, see Fig.~\ref{fig:clas-dvcs}. The leading amplitudes of these two asymmetries provide some sensitivity to the imaginary resp.~real part of CFF $\widetilde{\mathcal{H}}$. Also shown in the same figure is a very recent and preliminary result of a cross-section measurement. This observable is sensitive to the real part of the DVCS scattering amplitude. The first global analyses of DVCS data have started (e.~g.~\cite{Guidal,Moutarde}). Their aim is to extract in a model-independent way the real and imaginary parts of CFFs. 

Asymmetries with respect to transverse target polarization measured in hard exclusive electroproduction of real photons or vector mesons are the only observables for a proton target that depend linearly on the GPD $E$ without suppression by the GPD $H$. The Ji sum rule for the total angular momentum of quarks \cite{Ji}, $J_{\mathrm{q}}=\frac{1}{2}\lim_{t\rightarrow 0}\int_{-1}^{1}\mathrm{d}x\;x\left[H^{\mathrm{q}}(x,\xi,t)+E^{\mathrm{q}}(x,\xi,t)\right]$, involves both quark GPDs $H^{\mathrm{q}}$ and $E^{\mathrm{q}}$. Therefore the measurement of observables sensitive to $H^{\mathrm{q}}$ and $E^{\mathrm{q}}$ allows in principle a constraint on the total angular momentum carried by u- and d-quarks in the nucleon, though strongly model-dependent. {\sc Hermes} presented preliminary results on the single-spin asymmetry in exclusive production of an $\omega$- and $\phi$-meson on a transversely polarized proton target \cite{Augustyniak}. For such light vector mesons, the asymmetry is sensitive to the interference between GPDs $H$ and $E$. For the $\omega$-meson ($[u\overline{u}+d\overline{d}]/\sqrt{2}$), the leading asymmetry amplitude is negative, as expected from model calculations, and larger than the asymmetry for the neutral $\rho$-meson ($[u\overline{u}-d\overline{d}]/\sqrt{2}$), with the explanation that the contributions by the GPD $E$ for u-quarks and the GPD $E$ for d-quarks do not cancel each other, entering with opposite sign: $A_{\mathrm{UT}}\propto\Im\mathrm{m}\left[(2E^{\mathrm{u}}-E^{\mathrm{d}})/(2H^{\mathrm{u}}-H^{\mathrm{d}})\right]$. The transverse target-spin asymmetry for the $\phi$-meson, a pure sea ($s\overline{s}$) object, is compatible with zero within the experimental uncertainties. This 
suggests that the GPD $E$ for both sea quarks and gluons 
might be negligible.

As stated in the Introduction, the most intriguing theoretical issues for GPDs concern on one side setting up a global fit in the same spirit  of the ones existing for PDFs and TMDs,  and on the other, defining precisely the many facets of their partonic interpretation \cite{GolLiu}.  Global fitting is still at an early stage (see however Ref.\cite{Guidal,Moutarde,Mueller}).  The analysis of deeply virtual exclusive experiments presents in fact an increased level of complication because GPDs are complex amplitudes
(four chiral even and four chiral odd) embedded in the CFFs. Another consequence of describing the handbag diagram in the $t$-channel is that the number of kinematical variables increases, so that GPDs are fully determined by four independent variables, vs. three for the soft matrix elements in SIDIS, and two for the PDFs. 
There were recent attempts to reduce the number of variables connecting CFFs and GPDs using dispersion relations \cite{Teryaev}. However, substantial obstacles are given by the presence of both $t$-dependent thresholds, and in the interpretation of GPDs in the ERBL region \cite{Goldstein,GolLiu}. An improved knowledge of GPDs in a wide kinematical range will allow us to fulfill the ultimate goal of validating Ji's sum rule. This goal is already at reach in lattice calculations as shown in \cite{Meyer}. In order to understand the nature of orbital angular momentum it is also useful to consider targets of different spin. In \cite{Taneja} a sum rule analogous to Ji's, but for the deuteron was presented.  
Hard exclusive reactions can also be used to extract the chiral-odd GPDs in exclusive $\pi^o$ production with a $\gamma_5$ pion coupling \cite{Goldstein}, and in the more elusive pion plus $\rho^0_T$ exclusive production  \cite{Wallon} (the gauge invariance of the latter is not straightforward). Connections between chiral odd GPDs and TMDs have been worked out in models such as the one presented in \cite{Zavada}. 
Finally, are polarization effects important and are the formalism and approaches so far developed {\it directly} applicable in LHC processes? This question was partially addressed in \cite{Diehl} where multi-parton interactions in hadron-hadron collisions were analyzed using a GPD-based formalism, and also in \cite{Klasen} where the effect of polarization d.o.f. in the hadroproduction of SUSY particles was explored. 

\section{Summary and Conclusions}
\label{sec:summary}

In summary, a wealth of spin results were shown at the DIS 2010 conference. On the nucleon helicity sector, better constraints on the gluon polarization became possible through recent measurements at {\sc Rhic}. The contribution of the gluon spin to the spin of the nucleon turns out to be very small to compatible with zero. The production of W-bosons allows for access to the sea quark polarization. A lot of new and final results were presented on the TMD and GPD sector. These two fashionable frameworks of distribution functions allow one to picture the nucleon in a holographic way. Measurements at different facilities complement each other with regard to the covered phase space and the measured observables. All this on one side opens the stage for an already achievable global analysis of data. On the other, the  
formalism so far developed for both unpolarized and polarized hard processes will contribute to the interpretation and extraction of observables in a wide kinematical regime, in particular in the experimental setups available at the LHC.


\end{document}